# Gaussian Beams in Near-Zero Transition Metamaterials

**Fatema Alali[1] and Natalia M. Litchinitser[1]**

[1]*Department of Electrical Engineering, University at Buffalo, The State University of New York, Buffalo, New York, 14260, USA*

*faalali@buffalo.edu*

**Abstract:** We theoretically and numerically investigate the phenomenon of resonant field enhancement of Gaussian beams in two types of transition metamaterials: with a positive-zero-negative index profile and with a positive-zero-positive index profile and demonstrate strongly localized resonant field enhancement in both cases. This study is likely to have applications in the fields of nano-optics, sub-wavelength imaging, and nano-fabrication and lays a foundation for the studies of more complex vector and vortex beam propagation in graded-index metamaterials.

**OCIS codes:** (160.3918) Metamaterials, (080.2710) Inhomogeneous optical media.

---

## 1. Introduction

The emergence of graded index metamaterials gave rise to a number of new and unique functionalities [1]. Recently, we investigated the fundamental question of how electromagnetic waves propagate in an important class of inhomogeneous metamaterials, with material properties gradually changing from positive to negative values ("transition metamaterials" or TMMs) [2-5]. We discovered strongly polarization sensitive anomalous field enhancement near the zero refractive index point under oblique incidence of the wave on a realistic, lossy TMM layer, potentially enabling a variety of applications in microwave, terahertz, and optical metamaterials, including subwavelength transmission and low-intensity nonlinear optical devices. To date, most of the studies of TMMs were limited to plane wave propagation in such materials [2-7]. However, strictly speaking, the perfect plane wave can only be found in textbooks. In laboratory experiments, optical beams are limited in extent either by the instrument system used to produce them or by that used to observe them. Therefore, in this paper, we theoretically and numerically investigate the propagation of Gaussian beams in two types of TMMs: with a positive-zero-negative index profile, as shown in Fig. 1(a), and with a positive-zero-positive index profile, as shown in Fig. 1(b).

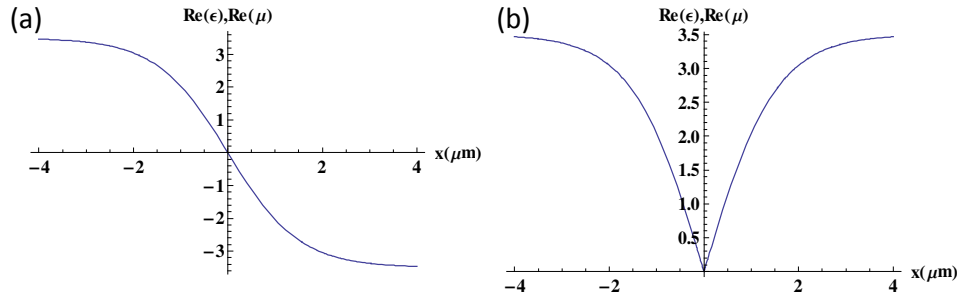

Figure 1. Real part of dielectric permittivity and magnetic permeability as functions of longitudinal coordinate for cases (a) positive-zero-negative transition metamaterial, and (b) positive-zero-positive transition metamaterial.

## 2. Theory

In many conventional optical problems, Gaussian beam propagation is often studied in a so-called paraxial approximation that requires that the beam size at the waist should be much larger than a wavelength. However, emerging fields of nano-optics and optical metamaterials (MMs), sub-wavelength imaging, and nano-fabrication necessitate the use of tightly focused beams producing highly localized field enhancements, in which case the paraxial approximation may break down [8-11]. Several authors addressed the problem of non-paraxial propagation of Gaussian beams in conventional media [12-15] and more recently in metamaterials [16]. Since this study focuses on novel MM-based structures enabling strongly localized electric as well as magnetic field enhancements, we investigated both paraxial and non-paraxial beam propagation in TMMs.
We will first briefly review angular spectrum representation of a Gaussian beam that involves expanding a complex wave field into a summation of an infinite number of plane waves. Consider a one-dimensional Gaussian beam propagating in x direction

$$E(x=0,y)=\frac{1}{\sqrt{2\pi}w_0}\exp\left(-\frac{y^2}{2w_0^2}\right),\qquad(1)$$

where $w_0$ is the half-width of the Gaussian beam waist.

The angular spectrum in the plane corresponding to $x=0$ is given by

$$\tilde{E}_{x=0}(qk)=\sqrt{2\pi}w_0\exp\left(-\frac{w_0^2 q^2 k^2}{2}\right),\qquad(2)$$

where $q$ is the direction cosine.

At a distance $x$, the field is given by

$$E(x,y)=\sqrt{2\pi}w_0\int_{-\infty}^{\infty}\exp\left(-\frac{w_0^2 q^2 k^2}{2}\right)\exp\left(ik\sqrt{1-q^2}\,x\right)\exp(ikqy)\,d(kq).\qquad(3)$$

Electromagnetic wave propagation in TMMs is described by the Helmholtz equation derived from Maxwell's equations for inhomogeneous mediums.

$$\frac{\partial^2 E_z}{\partial x^2}+\frac{\partial^2 E_z}{\partial y^2}-\frac{1}{\mu(x)}\frac{\partial \mu}{\partial x}\frac{\partial E_z}{\partial x}+\varepsilon(x)\mu(x)k^2 E_z=0,\qquad(4)$$

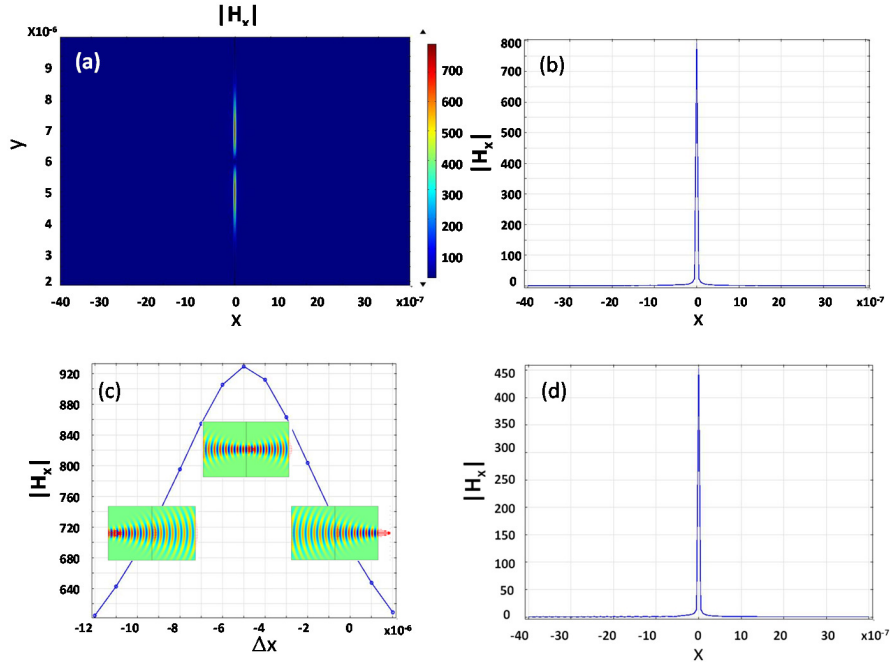

Figure 2. (a) Enhancement of Gaussian beam for the case of normal incidence, (b) Resonant enhancement in divergent region (Δx=−8μm see (c)), (c) The resonant enhancement as a function of Δx for the case of normal incidence of Gaussian beam with a waist of 1μm. The inserts show schematics of electric field distributions. (d) Resonant enhancement in the case of PZP.

where $k$ is the wave number in free space. It is noteworthy that the Gaussian beam in its standard form corresponds to the case of paraxial approximation and is not an exact solution of Maxwell's equations in free space. Therefore, we implemented a plane wave expansion, which yields an accurate description of the Gaussian beam as described above [16] and performed detailed numerical simulations in COMSOL Multiphysics aimed at systematically studying the Gaussian beam propagation near the zero refractive index point for the two cases shown in Fig. 1.

## 3. Numerical Results and Discussion

In this study, we considered a linearly polarized Gaussian beam propagating in the direction of the gradient of refractive index. We considered in detail a particular case of a real part of the refractive index $n$ (as well as $\varepsilon$ and $\mu$) gradually changing from positive to zero to negative values, a so-called positive-zero-negative metamaterial (PZN). Then, we compared field enhancement in the case of PZN shown in Fig. 2(a,b) with the case when the real part of refractive index $n$ gradually changed from positive to zero back to positive values, a so-called

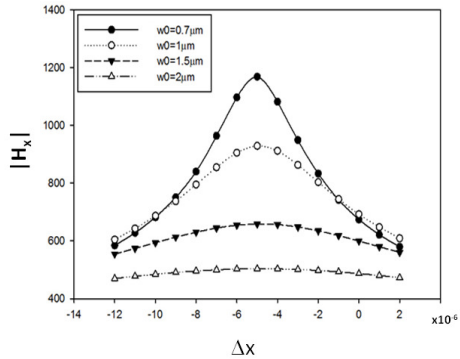

Figure 3. The resonant enhancement as a function of $\Delta x$ for different beam waists.

positive-zero-positive metamaterial (PZP) shown in Fig. 2(d).

Recall that in the case of a plane wave, anomalous field enhancement was predicted only at oblique incidence, while no enhancement occurred at normal incidence. This phenomenon can be understood by noticing that only at the oblique incidence electric (for the case of TM wave) or magnetic (for the case of TE wave) component of the electromagnetic wave has a projection in the direction of propagation. This longitudinal component gets enhanced as dielectric permittivity or magnetic permeability tends to zero. We showed that unlike the case of a plain wave, resonant enhancement for a Gaussian beam takes place near the zero index point even in the case of normal incidence, as shown in Fig. 2(a,b). This can be explained by noticing that the Gaussian beam is composed of a combination of plane waves, each propagating with a different propagation vector. We found that the maximum enhancement occurs near the point corresponding to the location of the waist of the beam in a homogenous medium with $\varepsilon = \mu = 3.5$ (Fig. 2c). Note that the waist of Gaussian beam in free space ($n=1$) is located at $\Delta x = 0$. In our model, however, the region at which the refractive index is homogenous has a value of $n=3.5$, which causes a shift in the waist location to $\Delta x = -5\mu m$. Figure 3 shows that the resonant enhancement of the longitudinal component of a Gaussian beam at normal incidence decreases with the beam waist $w_0$. This can be explained by the fact that the wider the waist of a Gaussian beam, the narrower its spectrum; therefore, its propagation resembles that of a plane wave.

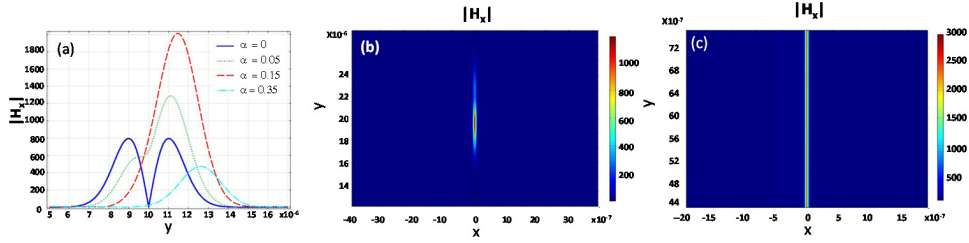

Figure 4. (a) Gaussian beam field enhancement along y-axis for different angles of incidence, (b) Enhancement of Gaussian beam at oblique incidence, (c) Enhancement of a plane wave at oblique incidence.

Although negative index metamaterials at optical wavelengths were demonstrated first in a form of a thin film and later in 3D case (with a thickness still of the order of a wavelength), it is still challenging to fabricate bulk structures containing transition layers with a refractive index changing from positive to negative values over a length greater or equal to the wavelength of the light beam. We investigated Gaussian beam propagation near the zero index point in positive-zero-positive metamaterials and demonstrated a possibility of strong field enhancement in this case as well (Fig. 2d). This case is of particular interest for potential application, as it does not rely of the availability of bulk negative index materials, which are still quite challenging to realize experimentally. These results suggest that instead, one could use positive-zero-positive index and realize strongly localized field enhancement. Detailed studies of plane wave propagation in this case will be reported elsewhere.

We performed a parametric study of the dependence of the field enhancement of the angle of incidence, as shown in Fig. 4(a). This figure clearly indicates that depending on the angle of incidence, different parts of the beam are enhanced. While in the case of normal incidence, Gaussian wings are more enhanced; in the oblique incidence case, the center of the beam is more enhanced. Figures 4(b,c) also show stronger field localization in the case of an oblique Gaussian beam as opposed to the case of plane wave enhancement.

Next, as we mentioned in the introduction, tightly focused beams are of significant interest for applications in optical trapping. In this context, various donut-shaped optical beams, including cylindrical vector beams and vortices are of particular interest. As a first step toward understanding of propagation of such beams in TMMs, we investigated evolution of the first-order ($n=1$) Hermite-Gaussian beam near zero-index transition. Before discussing the results of numerical simulations, let us highlight the difference between these two beams in terms of angular distribution of angular components as shown in Fig. 5. In the case of a conventional Gaussian beam ($n=0$), peak intensity corresponds to the central (small angle) components of the beam. Assuming that we consider the case of normal incidence, even if the beam is tightly focused, these k-vector components maintain a nearly perpendicular position to the plane of $n = 0$, which implies that there are no components of either electric or magnetic field in the direction of propagation, and as a result, no resonant field enhancement on the optical axis (where the intensity of the beam is maximized) [2]. On the other hand, the longitudinal component of the field of the Hermite-Gaussian beam has a maximum at the beam focus with much larger field strength. Indeed, this strong longitudinal component of the Hermite-Gaussian beam was proposed to be used to accelerate charged particles along the beam axis in linear particle accelerators [16] or to image the spatial orientation of molecular transition dipoles [17,18]. Figure 6 shows a comparison of the field enhancement for the cases of Gaussian and Hermite-Gaussian beams, indicating a possibility that the enhancement of oblique wave vector components (q≠0) can be seen in response to a Hermite-Gaussian beam. We note that in part of the study, for simplicity, we used the paraxial approximation for both the fundamental Gaussian beam and Hermite-Gaussian beam although it can be generalized by using non-paraxial solution for the Hermite-Gaussian beam that is available in the literature [19]. Figure 6 confirms our qualitative prediction of the presence of large field enhancement for the axial longitudinal components of the field in the case of the Hermite-Gaussian beam (Fig. 6b) that is absent in the case of a fundamental Gaussian beam (Fig. 6a).

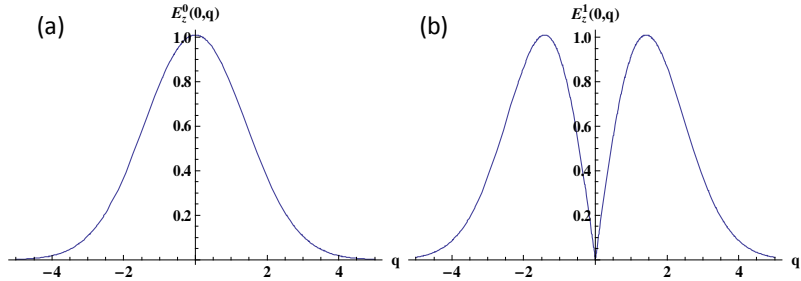

Figure 5. Fourier Transform of (a) a fundamental Gaussian beam, (b) the first order Hermite-Gaussian beam.

Finally, it is well-known that a majority of demonstrated optical metamaterials are lossy structures. While in our study, we only included small losses near the zero index point (where even the very small imaginary part of $n$ cannot be neglected as it is not small when compared to its real part); here, we briefly consider the case of increased losses as shown in Fig. 7(a). Figure 7(b) confirms that field enhancement still occurs near the zero refractive index point

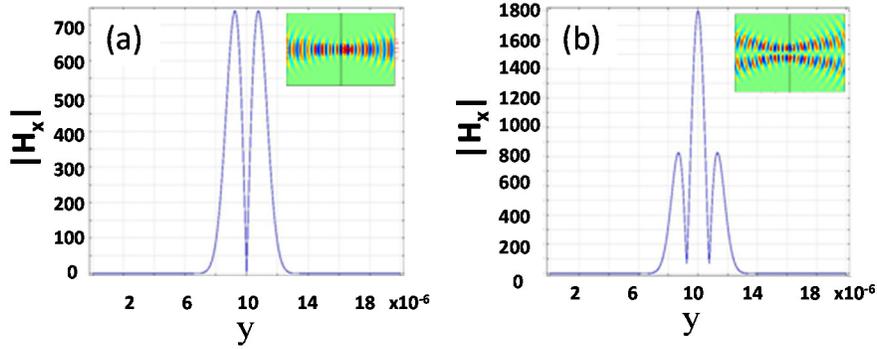

Figure 6. A comparison of resonant field enhancement at normal incidence in the case of (a) fundamental Gaussian beam, and (b) Hermite-Gaussian beam.

even though it is the region with the highest losses.

## 4. Summary

We investigated the phenomenon of resonant field enhancement of Gaussian beam propagation in TMMs. Previously, we demonstrated that the field enhancement of plane waves in TMMs is strongly dependent on the angle of incidence and cannot take place in the case of normal incidence (as there is no longitudinal component of either electric or magnetic field in this case). Based on these initial studies herein, we predicted and demonstrated that in the case of Gaussian beam resonant enhancement occurs for both normal and oblique incidence due to the fact that the beam is composed of an infinite sum (integral) of plane waves, each with different angular vectors. We established that maximum enhancement occurs in the waist region because different plane wave components exhibit no phase delay with respect to each other in the waist plane. Resonant enhancement was found in both the positive-zero-negative index profile and the positive-zero-positive index profile. The latter case may be of a special interest from a practical viewpoint. We confirmed that the field enhancement can be achieved even in a more realistic case of lossy metamaterials. Finally, we performed preliminary studies of the propagation of Hermite-Gaussian beams in TMMs and demonstrated important differences between them with respect to fundamental Gaussian beam enhancement. This study lays a foundation for the studies of more complex vector and vortex beam propagation in transition metamaterials.

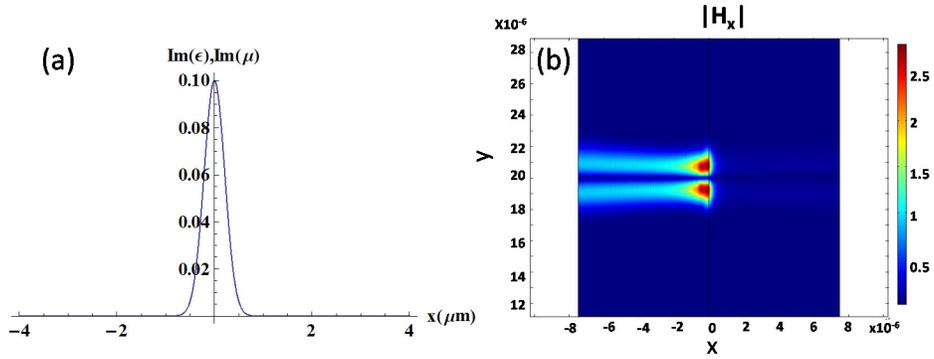

Figure 7. (a) Losses profile in TMM, (b) Corresponding field enhancement.

The authors would like to thank Dr. Kildishev and Dr. Mozjerin for their help at the early stage of this work. This research was supported by the US Army Research Office under award # W911NF-11-1-0333.